\documentclass[prb,aps,twocolumn,showpacs,nofootinbib,superscriptaddress]{revtex4}

\usepackage{graphicx}
\usepackage{stmaryrd}
\usepackage{rotating}
\usepackage{amsmath}
\usepackage{amsfonts}
\usepackage{amssymb}
\usepackage[countmax]{subfloat}
\usepackage{dcolumn} 
\usepackage{bm} 
\usepackage{color}

\usepackage{float}
\usepackage{latexsym}
\usepackage{wasysym}

\begin{document}

\title{Scaling Theory of ${\mathbb Z}_{2}$ Topological Invariants}

\author{Wei Chen} 

\affiliation{Theoretische Physik, ETH-Z\"urich, CH-8093 Z\"urich, Switzerland}

\author{Manfred Sigrist} 

\affiliation{Theoretische Physik, ETH-Z\"urich, CH-8093 Z\"urich, Switzerland}

\author{Andreas P. Schnyder} 

\affiliation{Max-Planck-Institut f$\ddot{u}$r Festk$\ddot{o}$rperforschung, Heisenbergstrasse 1, D-70569 Stuttgart, Germany}

\date{\rm\today}

\begin{abstract}

{For inversion-symmetric topological insulators and superconductors characterized by ${\mathbb Z}_{2}$ topological invariants, two scaling schemes are proposed to judge topological phase transitions driven by an energy parameter. The scaling schemes  renormalize either the phase gradient or the second derivative of the Pfaffian of the time-reversal operator, through which the renormalization group flow of the driving energy parameter can be obtained. The Pfaffian near the time-reversal invariant momentum is revealed to display a universal critical behavior for a great variety of models examined. 
 }

\end{abstract}






\maketitle

\section{introduction}

An issue that is of fundamental significance in the field of topological order is how to drive a system into topologically nontrivial states, such that their intriguing properties can be exploited. The topologically nontrivial state usually requires inverting two bulk bands of different symmetries, which is typically achieved by tuning an energy parameter such as chemical potential\cite{Kitaev01,Lutchyn10,Oreg10,Mourik12}, hopping amplitude\cite{Su79}, external magnetic field\cite{Mourik12}, interface coupling\cite{Choy11,NadjPerge13,Braunecker13,Pientka13,Klinovaja13,Vazifeh13,Rontynen14,Kim14,Sedlmayr15,Chen15}, etc. Consequently, to identify the topological phase transitions, the usual way is to seek the value of the tuning energy parameter at which the gap of the two bulk bands closes. 
Provided the system belongs to a suitable symmetry class\cite{Schnyder08,Kitaev09,Chiu15}, one 
then expects that the band inversion leads to a topologically nontrivial phase.


Recently, another recipe for judging topological phase transitions has been developed, namely the scaling theory of
topological invariants\cite{Chen16}. Drawing an analogy from Kadanoff's scaling approach\cite{Kadanoff66}, which lays the foundation for the renormalization group (RG)  in Landau's order parameter paradigm, the issue of scaling is itself of importance. 
Since topologically nontrivial systems do not necessarily possess order parameters,
the very concept of scaling must follow a different principle than in Kadanoff's approach.
Indeed, it was shown in Ref.~\onlinecite{Chen16}, that for inversion-symmetric topologically nontrivial systems
the meaning of scaling corresponds to  stretching a knotted string, thereby  revealing the number of knots it contains\cite{Chen16}. 
This is in contrast to Kadanoff's approach, where short range degrees of freedom are successively integrated out.


The philosophy behind the scaling theory of topological invariants is the observation that topological properties are characterized by a winding number that is calculated from a momentum-space integration of a certain curvature function, e.g.,  a Berry curvature\cite{Berry84,Xiao10,Thouless82}, a Berry connection\cite{Zak89}, or a winding number density\cite{Ryu10}. Although the curvature function (local curvature of a closed string) changes as the tuning energy parameter varies, the winding number (number of knots) remains the same, as long as the system stays in the same topological phase. The scaling procedure is a process that converges the curvature function into its fixed point configuration without changing the winding number.
Hence, the ``RG flow" of the tuning energy parameter is
akin to stretching a string until all the knots are tightened.

The purpose of this paper is to derive the scaling theory for 
 inversion-symmetric  topological systems characterized by ${\mathbb Z}_{2}$ topological invariants\cite{Schnyder08,Kitaev09,Chiu15}.
In contrast to ${\mathbb Z}$ invariants, which are computed from a Berry curvature (or a winding number density), 
the ${\mathbb Z}_{2}$ invariants can be expressed in terms of the matrix elements of the time-reversal operator $\Theta$\cite{Kane05,Moore07,Fu07,Fu07_2},
\begin{eqnarray}
m_{\alpha\beta}({\bf k},M)=\langle e_{\alpha}({\bf k},M)|\Theta| e_{\beta}({\bf k},M)\rangle ,
\label{time_reversal_operator_matrix}
\end{eqnarray}
where $|e_{\alpha}({\bf k},M)\rangle$ is the $\alpha$-th occupied eigenstate at momentum ${\bf k}$. The tuning energy parameter is denoted by $M$ (previously denoted by $\Gamma$ in Ref.\onlinecite{Chen16}). Note that  ${\mathbb Z}_2$ invariants can also be defined in terms of the sewing matrix\cite{Fu06} $w_{\alpha\beta}({\bf k},M)=\langle e_{\alpha}(-{\bf k},M)|\Theta| e_{\beta}({\bf k},M)\rangle$.
(For alternative definitions of the ${\mathbb Z}_{2}$ invariants see also Ref.~\onlinecite{Chiu15}.)
Here, however we focus on ${\mathbb Z}_{2}$ invariants given in terms of Eq.~(\ref{time_reversal_operator_matrix}), 
because $m_{\alpha\beta}$ is antisymmetric at any ${\bf k}$ and hence its Pfaffian is a continuous function that can be rescaled to judge topological phase transitions.
The Pfaffian of $w_{\alpha\beta}$, on the other hand, is only defined at the discrete time-reversal invariant momenta, for which we do not see how a continuous scaling scheme can be constructed. 


There are two possible ways to obtain the ${\mathbb Z}_{2}$ invariant from Eq.~(\ref{time_reversal_operator_matrix}): (i) from the phase of  ${\rm Pf}(m)$ or (ii) from the sign of  ${\rm Pf} ( m )$. 
In the following we demonstrate that 
for each of these two possibilities there
exists a continuous scaling scheme, which employs  either the phase gradient or second derivative of ${\rm Pf} ( m )$.
Under these scaling schemes, the RG flow of any tuning energy parameter can be obtained, from which the topological phase transitions can be identified, regardless of the symmetry class of the system. Furthermore, we obtain a length scale defined from the divergence of the phase gradient of ${\rm Pf} ( m )$ [or from the second derivative of ${\rm Pf} ( m )$] at time-reversal invariant momenta. This length scale signals scale invariance at the fixed points and critical points, thereby implying universal critical behavior.



\section{Scaling scheme using phase of the Pfaffian}

Let us start by considering the scaling scheme which employs the phase of ${\rm Pf} ( m )$. 
Before we examine several examples, we first discuss the general formalism of this scaling scheme.

\subsection{General formalism}

\begin{figure}[t!]
\begin{center}
\includegraphics[clip=true,width=0.99\columnwidth]{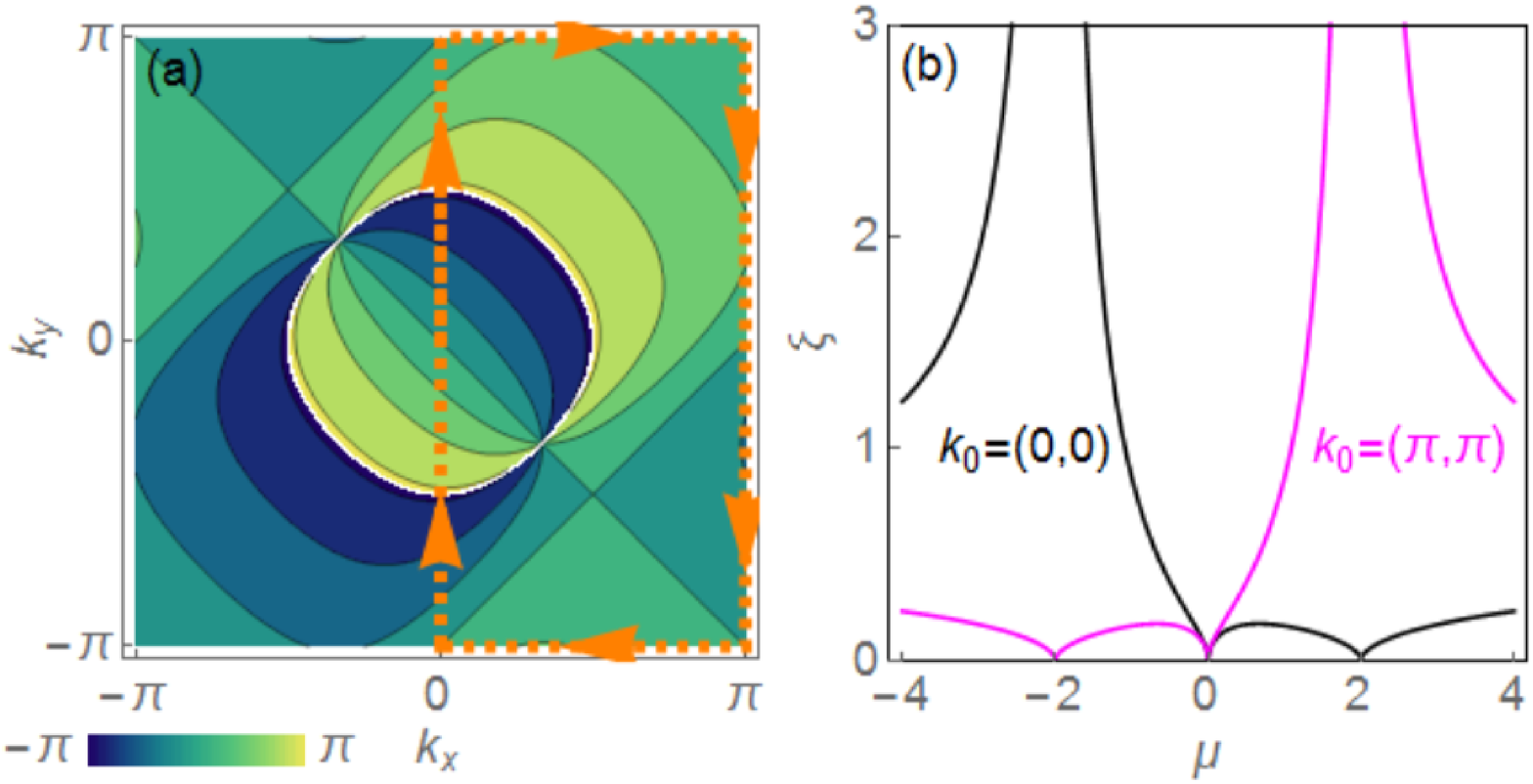}
\includegraphics[clip=true,width=0.9\columnwidth]{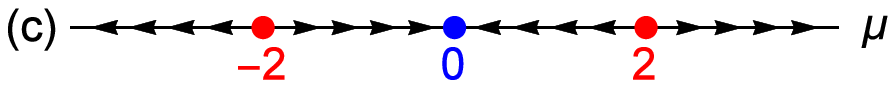}
\caption{(color online) (a) The phase $\varphi({\bf k},\mu)$ of the Pfaffian ${\rm Pf}(m)$ 
for the graphene model given by Eq.~(\ref{Carpentier_minimal_model}).
Here, we set $\mu=-1$, in which case $\varphi({\bf k},\mu)$  is an odd function along the diagonal direction ${\hat{\bf k}}_{s}=\left(1/\sqrt{2},1/\sqrt{2}\right)$.  Hence, the gradient of $\varphi({\bf k},\mu)$ is even along this direction.
The parameter choice $(\mu, t)=(-1, +1)$ corresponds to a topologically nontrivial state with two vortices in $\varphi({\bf k},\mu)$, i.e., one vortex in each half BZ. The dashed line indicates the integration contour of Eq.~(\ref{Z2_index}). With this integration contour
Eq.~\eqref{Z2_index}  counts the number of vortices in the half BZ with $k_x > 0$. Intuitively, the scaling procedure pushes the two vortices away from ${\bf k}_{0}=(0,0)$ or $(\pi,\pi)$ at which the gap closes. Hence, the system gradually flows away from the critical point $M_{c}$ under the scaling procedure. (b) The length scale $\xi$ and (c) the RG flow described by Eq.~(\ref{Carpentier_model_RG_equation}), which is the same for the two choices of ${\bf k}_{0}$. Red dots denote the critical points while the blue dot is the fixed point.}  
\label{fig:Carpentier_model}
\end{center}
\end{figure}

One possibility to obtain a ${\mathbb Z}_{2}$ invariant from ${\rm Pf}(m)$ is by use of the following formula\cite{Kane05}
\begin{eqnarray}
\nu=\frac{1}{2\pi i}\oint_{\partial \frac{1}{2}BZ}d\log\left[{\rm Pf}(m)\right]\;\;\;{\rm mod}\;2\;,
\label{Z2_index}
\end{eqnarray}
which simply counts the number of windings of the phase
\begin{eqnarray}
\varphi({\bf k},M)=\arg\left[{\rm Pf}(m)\right]\; 
\label{phase_of_Pfaffian}
\end{eqnarray}
 in the half Brillouin zone (BZ) $ \frac{1}{2}BZ$ in 2D\cite{Fruchart13} (see Fig.~\ref{fig:Carpentier_model}(a) for an example), or the number of vortices in the half BZ of a high-symmetry plane in a 3D model. That is, Eq.~(\ref{Z2_index})  represents the ${\mathbb Z}_{2}$ invariant of 2D systems with time-reversal symmetry (class AII), as well as the weak topological invariants in 3D, which are
  defined for each high-symmetry plane $k_{x}=0$, $k_{y}=0$, and $k_{z}=0$. 
Due to time-reversal symmetry, the vortices in $\varphi({\bf k},M)$  appear as  pairs, one in each half BZ. States with an
odd number of vortices in $\frac{1}{2}BZ$   are topologically distinct from those with an even number of vortices
in $\frac{1}{2}BZ$.


The scaling scheme is constructed in the following way. We choose a scaling direction ${\hat{\bf k}}_{s}$ to define the scaling function  
\begin{eqnarray}
F({\bf k},M)={\hat{\bf k}}_{s}\cdot{\boldsymbol\nabla}_{\bf k}\varphi({\bf k},M)\;,
\label{FkGamma_definition}
\end{eqnarray}
which is an even function around a time-reversal invariant momentum ${\bf k}_{0}$
\begin{eqnarray}
F({\bf k}_{0}+\delta k{\hat{\bf k}}_{s},M)=F({\bf k}_{0}-\delta k{\hat{\bf k}}_{s},M)\;.
\end{eqnarray}
Given an initial value of $M$, we seek for a new $M^{\prime}$ that satisfies
\begin{eqnarray}
F({\bf k}_{0},M^{\prime})=F({\bf k}_{0}+\delta k{\hat{\bf k}}_{s},M)\;,
\label{scaling_procedure_general}
\end{eqnarray}
where $\delta k{\hat{\bf k}}_{s}$ is a small displacement along the scaling direction. The mapping $M\rightarrow M^{\prime}$ gives an RG flow that identifies the topological phase transition. The reason for considering Eq.~(\ref{scaling_procedure_general}) is the deviation-reduction mechanism detailed in Ref.~\onlinecite{Chen16}. In short, we write $F({\bf k},M)$ as a sum of the fixed point configuration $F_{f}$ and the deviation $F_{v}$ away from it
\begin{eqnarray}
F({\bf k},M)=F_{f}({\bf k},M_{f})+F_{v}({\bf k},M)\;.
\label{F_Ff_Fv}
\end{eqnarray}
Because the deviation part $F_{v}({\bf k},M)$ in inversion symmetric systems is an even function of ${\bf k}$ and must integrate to zero to conserve the ${\mathbb Z}_{2}$ invariant, it can be expanded in terms of Fourier cosine series.
One can then rigorously proved that the amplitude of the deviation part at the high symmetry point $F_{v}({\bf k}_{0},M)$ gradually reduces to zero under the operation of Eq.~(\ref{scaling_procedure_general}). Hence, $F({\bf k},M)$ gradually converges to the fixed point configuration $F_{f}({\bf k},M_{f})$ that is invariant under this scaling procedure. As a result, the system gradually flows away from the critical point $M_{c}$, where $F({\bf k}_{0},M)$ diverges, to the fixed point $M_{f}$ where $F({\bf k}_{0},M)$ flattens to second order.

We remark that similar to all other RG approaches, this scaling scheme finds the fixed point $M_{f}$ but it does not give a physical meaning to $M_{f}$, i.e., it does not tell us whether a particular $M_{f}$ is topologically trivial or nontrivial, which nevertheless can be clarified by directly calculating the ${\mathbb Z}_{2}$ invariant at $M_{f}$ or at any $M$ that flows to this $M_{f}$.

Expanding Eq.~(\ref{scaling_procedure_general}) in both $dM=M^{\prime}-M$ and $dl=\delta k^{2}$ yields the leading order RG equation\cite{Chen16}
\begin{eqnarray}
\frac{dM}{dl}=\frac{1}{2}\frac{({\hat{\bf k}_{s}}\cdot\nabla_{\bf k})^{2}F({\bf k},M)|_{{\bf k}={\bf k}_{0}}}{\partial_{M}F({\bf k}_{0},M)}\;.
\label{generic_RG_equation}
\end{eqnarray}
Since there is only one energy parameter $M$, one can always map the RG equation into the equation of motion $dM/dl=-\partial V/\partial M$ of an overdamped particle in a conservative potential, where $M$, $l$, and $V$ play the roles of coordinate, time, and potential energy, respectively\cite{Chen04,Chang05}. The extremal of $F({\bf k},M)$ at ${\bf k}_{0}$ suggests that near ${\bf k}_{0}$, one can expand\cite{Chen16} 
\begin{eqnarray}
F({\bf k}_{0}+{\hat{\bf k}_{s}}\delta k,M)=\frac{F({\bf k}_{0},M)}{1\pm\xi^{2}\delta k^{2}}\;.
\label{correlation_length_definition}
\end{eqnarray} 
The length scale $\xi$ characterizes the scale-invariance at the fixed point and critical points, as we shall see in the examples below.

For $4\times 4$ Dirac Hamiltonians that include sublattice or orbital degrees of freedom, our discussion follows 
along the lines of Ref.~\onlinecite{Fu07}, where ${\mathbb Z}_{2}$ invariants for inversion symmetric systems have been considered. A generic Dirac Hamiltonian can be written in terms of the Gamma matrices $\Gamma_{a}$ as
\begin{eqnarray}
H({\bf k},M)=d_{0}({\bf k})I+\sum_{a=1}^{5}d_{a}({\bf k},M)\Gamma_{a} ,
\label{Dirac_Hamiltonian}
\end{eqnarray}
where the corresponding spinor basis consists of a sublattice degree of freedom ($A/B$) and a spin degree of freedom
($\uparrow / \downarrow$) with
\begin{eqnarray}
\left(A,B\right)\otimes\left(\uparrow,\downarrow\right)=\left(A\uparrow,A\downarrow,B\uparrow,B\downarrow\right)\; .
\end{eqnarray}
For the Gamma matrices $\Gamma_a$ we adopt the representation\cite{Fu07} 
\begin{eqnarray}
\Gamma_{a}=\left\{\sigma_{x}\otimes I,\sigma_{y}\otimes I,\sigma_{z}\otimes s_{x},\sigma_{z}\otimes s_{y},\sigma_{z}\otimes s_{z}\right\}\;,
\label{Gamma_matrix_Fu_Kane}
\end{eqnarray}
where $\sigma_{i}$ acts on sublattice space and $s_{i}$ acts on spin space, and $I$ is the identity matrix. With these definitions the time-reversal 
operator is given by 
\begin{eqnarray}
\Theta=i\left(I\otimes s_{y}\right)K\;,
\end{eqnarray}
where $K$ is the complex conjugation operator. The parity operation is given by
\begin{eqnarray}
P=P^{-1}=\sigma_{x}\otimes I=\Gamma_{1}\; 
\end{eqnarray}
and therefore inverts the two sublattices.
The $\Gamma_{a}$ matrices transform  under these operations as
\begin{eqnarray}
\Theta \Gamma_{a}\Theta^{-1}=P\Gamma_{a}P^{-1}=
\left\{
\begin{array}{ll}
+\Gamma_{a} & {\rm if}\;a=1 \\
-\Gamma_{a} & {\rm if}\;a\neq 1 
\end{array}
\right. \; .
\end{eqnarray}
Thus, $\Gamma_{a}$ is even under the combined symmetry $P\Theta$.


After diagonalization, the energy spectrum is obtained $E_{\pm}=d_{0}\pm d$ with $d=\sqrt{\sum_{a=1}^{5}d_{a}^{2}}$. One can label the four eigenstates by $|e_{\alpha}({\bf k},M)\rangle$, and calculate the time-reversal operator matrix elements $m_{\alpha\beta}$ accordingly. The corresponding Pfaffian is
\begin{eqnarray}
{\rm Pf}(m)=-\frac{d_{1}\left(d_{1}-id_{2}\right)}{d\sqrt{d^{2}-d_{5}^{2}}}\; .
\label{Pfaffian_generic_form}
\end{eqnarray}
Hence, the gradient of the phase of the Pfaffian is only determined by $d_{1}$ and $d_{2}$ 
\begin{eqnarray}
\nabla_{\bf k}\varphi({\bf k},M)=-\nabla_{\bf k}\arctan\left(\frac{d_{2}({\bf k},M)}{d_{1}({\bf k},M)}\right)\;.
\label{gradient_of_phase_generic_form}
\end{eqnarray}
The RG procedure, Eqs.~(\ref{FkGamma_definition}) and (\ref{scaling_procedure_general}), can then be applied, with a proper choice of scaling direction ${\hat{\bf k}}_{s}$ which depends on the precise form of $d_{1}$ and $d_{2}$, as shown in the examples below. 



\subsection{Graphene model}

Our first example concerns the graphene model of Fu and Kane\cite{Fu07,Fruchart13}. We choose to parametrize it by dimensionless chemical potential $\mu$ and hopping amplitude $t$, where the hopping amplitude is set to be the energy unit $t=1$. The model is a 2D $4\times 4$ Dirac Hamiltonian of class AII with
\begin{eqnarray}
d_{1}({\bf k},\mu)&=&\mu+\cos k_{x}+\cos k_{y}\;,
\nonumber \\
d_{2}({\bf k},\mu)&=&\sin k_{x}+\sin k_{y}\;,
\nonumber \\
d_{5}({\bf k},\mu)&=&\sin k_{x}-\sin k_{y}-\sin(k_{x}-k_{y})\; ,
\label{Carpentier_minimal_model}
\end{eqnarray}
where the topological phase transition is driven by varying $\mu$. The field $\varphi({\bf k},\mu)$ is shown in Fig.~\ref{fig:Carpentier_model} (see also Ref.~\onlinecite{Fruchart13}). One can clearly see that $\varphi({\bf k},\mu)$ is odd along the diagonal direction ${\hat{\bf k}}_{s}=(1/\sqrt{2},1/\sqrt{2})$, therefore its gradient $F({\bf k},\mu)$ is even along this direction around both ${\bf k}_{0}=(0,0)$ and ${\bf k}_{0}=(\pi,\pi)$ of the BZ. From Eqs.~(\ref{Pfaffian_generic_form}) and (\ref{gradient_of_phase_generic_form}), we choose the scaling function to be
\begin{eqnarray}
&&F({\bf k},\mu)=\frac{1}{\sqrt{2}}\left(\partial_{k_{x}}+\partial_{k_{y}}\right)\varphi({\bf k},\mu)
\nonumber \\
&&=-\frac{1}{\sqrt{2}}\left(\partial_{k_{x}}+\partial_{k_{y}}\right)\arctan\left(\frac{d_{2}}{d_{1}}\right)
\nonumber \\
&&=-\frac{1}{\sqrt{2}}\frac{\left(d_{1}-\mu\right)d_{1}+d_{2}^{2}}{\left(d_{1}^{2}+d_{2}^{2}\right)} .
\end{eqnarray}
Using Eq.~(\ref{scaling_procedure_general}) with the small displacement $\delta k{\hat{\bf k}}_{s}=\delta k(1,1)/\sqrt{2}$ and writing $d\mu=\mu^{\prime}-\mu$, $dl=\delta k^{2}$, one obtains the leading order RG equation 
\begin{eqnarray}
\frac{d\mu}{dl}&=&\mu\left(\frac{1}{4}\mp\frac{1}{\mu\pm 2}\right)=\beta_{gra}(\mu)\;,
\nonumber \\
\xi&=&\left|\frac{\beta_{gra}(\mu)}{\mu\pm 2}\right|^{1/2}\; .
\label{Carpentier_model_RG_equation}
\end{eqnarray}
The upper and lower signs in Eq.~\eqref{Carpentier_model_RG_equation} correspond to   the choice ${\bf k}_{0}=(0,0)$ and ${\bf k}_{0}=(\pi,\pi)$, respectively. The RG equation correctly captures the critical point $\mu_{c}=\pm 2$. The fixed points are at $\mu_{f}=\left\{0,\pm\infty\right\}$. As shown in Fig.~\ref{fig:Carpentier_model}, the length scale $\xi$ diverges at the critical point $\mu_{c}$ when ${\bf k}_{0}$ is chosen to be the gap-closing momentum at this $\mu_{c}$. For instance, $\xi$ diverges at $\mu_{c}=2$ when one chooses ${\bf k}_{0}=(\pi,\pi)$ which is the gap-closing momentum at this $\mu_{c}$. Thus the divergence of $\xi$ reflects the divergence of $F({\bf k}_{0},\mu)$ at the gap-closing ${\bf k}_{0}$ when $\mu$ approaches $\mu_{c}$.

\begin{figure}[t!]
\begin{center}
\includegraphics[clip=true,width=0.7\columnwidth]{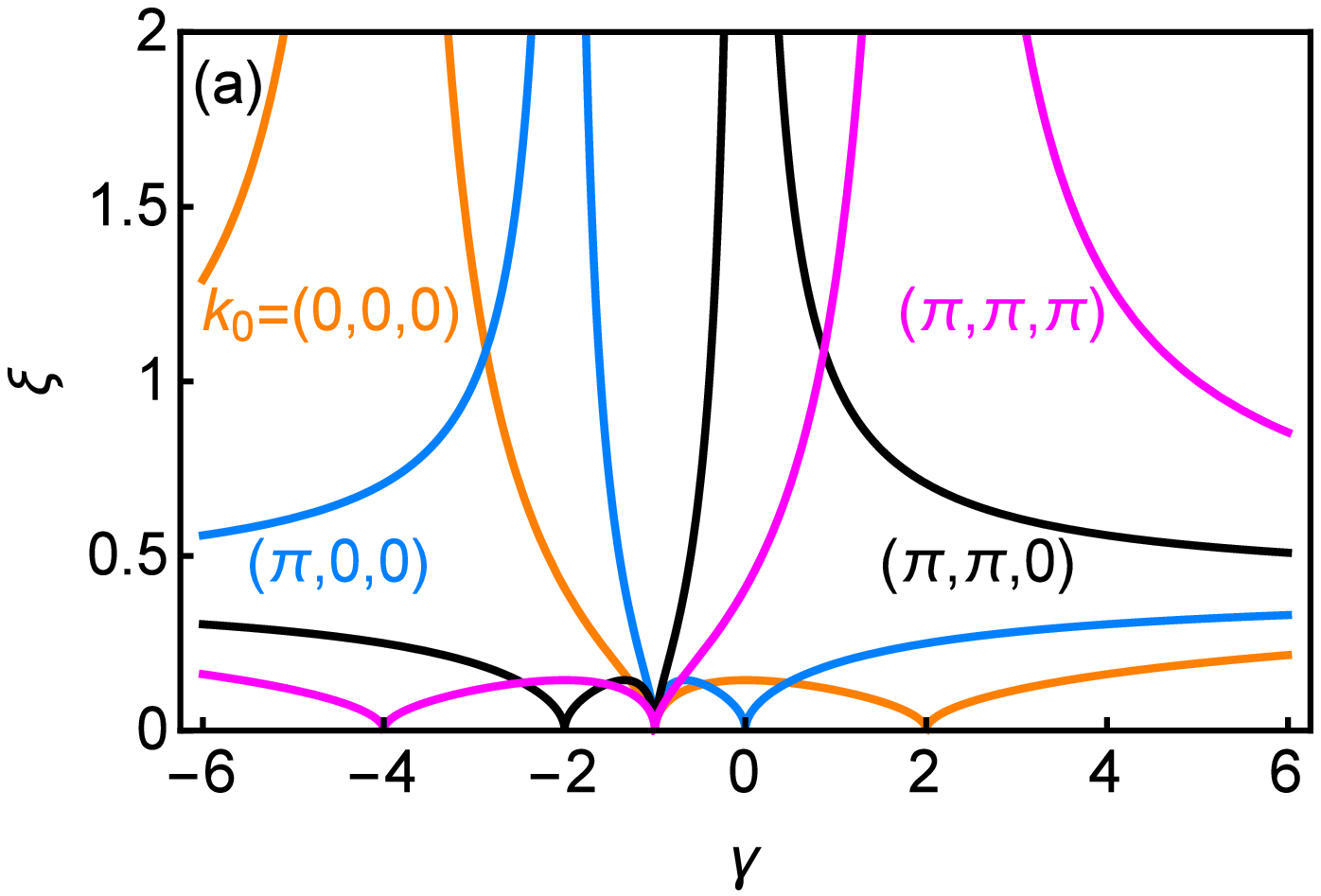}
\includegraphics[clip=true,width=0.8\columnwidth]{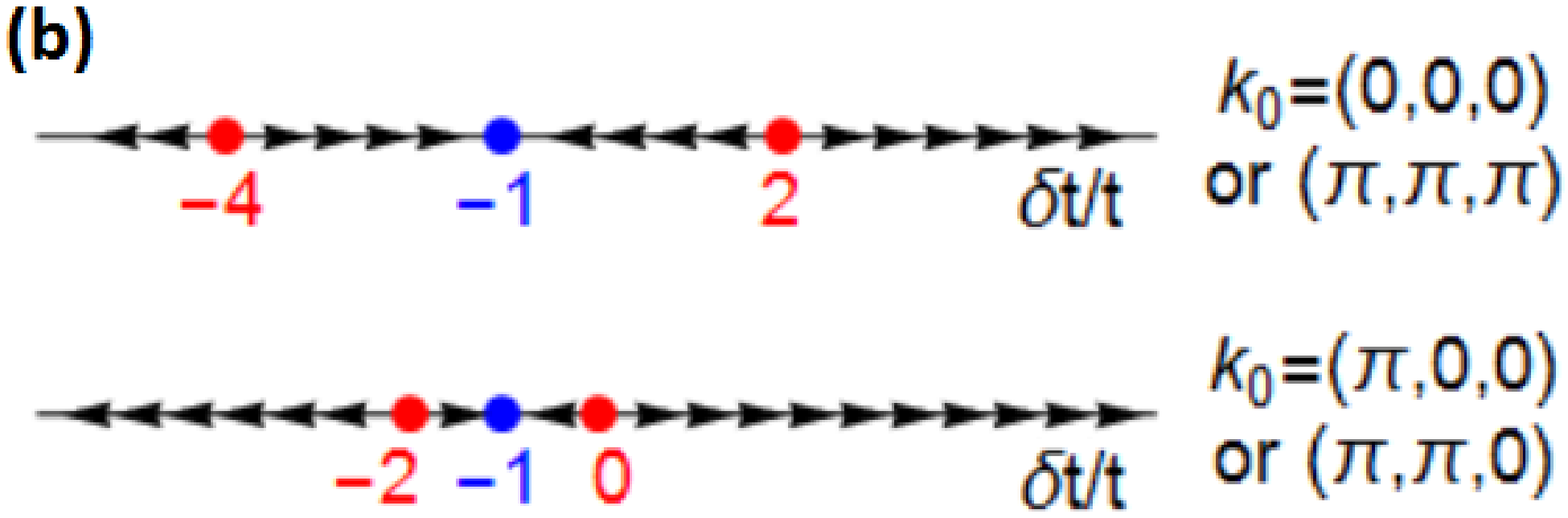}
\caption{(color online) (a) The length scale $\xi$ in the diamond lattice model described by Eq.~(\ref{RG_eq_diamond_lattice}), with the four choices of ${\bf k}_{0}$. Each choice represents a set of $\left\{N_{0},N_{\pi}\right\}$. (b) The RG flow of $\delta t/t$ with the four choices of ${\bf k}_{0}$. Red and blue dots denote the critical points and fixed points, respectively. 
} 
\label{fig:diamond_lattice_model}
\end{center}
\end{figure}

\subsection{Diamond lattice model}

Next, we discuss the 3D diamond lattice model of Ref~\onlinecite{Fu07}. 
This 3D model belongs to symmetry class AII and is described by Eq.~\eqref{Dirac_Hamiltonian}
with the ${\bf d}({\bf k},\delta t)$ given by
\begin{eqnarray} \label{3Ddiamond_lattice_model}
d_{1}({\bf k},\delta t)&=&t+\delta t+t\left(\cos k_{x}+\cos k_{y}+\cos k_{z}\right)\;,
\nonumber \\
d_{2}({\bf k},\delta t)&=&t\left(\sin k_{x}+\sin k_{y}+\sin k_{z}\right)\;,
\nonumber \\
d_{3}({\bf k},\delta t)&=&\lambda_{SO}\left[\sin k_{y}-\sin k_{z}\right.
\nonumber \\
&&\left.-\sin\left(k_{y}-k_{x}\right)+\sin\left(k_{z}-k_{x}\right)\right]\;,
\nonumber \\
d_{4}({\bf k},\delta t)&=&\lambda_{SO}\left[\sin k_{z}-\sin k_{x}\right.
\nonumber \\
&&\left.-\sin\left(k_{z}-k_{y}\right)+\sin\left(k_{x}-k_{y}\right)\right]\;,
\nonumber \\
d_{5}({\bf k},\delta t)&=&\lambda_{SO}\left[\sin k_{x}-\sin k_{y}\right.
\nonumber \\
&&\left.-\sin\left(k_{x}-k_{z}\right)+\sin\left(k_{y}-k_{z}\right)\right]\; .
\end{eqnarray}
We seek for the weak topological phase transitions of this model, which are driven by $\delta t$. That is, we look for changes in the weak topological invariants that are defined for the three high-symmetry planes,
$k_x =0$, $k_y=0$, and $k_z=0$,
of the 3D BZ.
These topological phase transitions can be conveniently captured by choosing the scaling direction ${\bf k}_{s}=(1,1,1)/\sqrt{3}$ that has nonzero projection in  all three high-symmetry planes.
With this choice for  ${\bf k}_{s}$ we apply  Eqs.~(\ref{FkGamma_definition}) and (\ref{scaling_procedure_general})
together with Eq.~(\ref{gradient_of_phase_generic_form}), which is an even function of ${\bf k}$.


The gap of the 3D diamond lattice model~\eqref{3Ddiamond_lattice_model} can close at any corner of the first quartet of the BZ, i.e., at ${\bf k}_{0}=(k_{0x},k_{0y},k_{0z})$ with $k_{0i}=0$ or $\pi$. Thus, to capture all the critical points in the whole $\delta t$ parameter space, it is necessary to perform the scaling procedure in every ${\bf k}_{0}$, similar to the situation of a topological insulator in $d$-dimensional cubic lattice\cite{Chen16}. For a given ${\bf k}_{0}$ in the first quartet of the BZ, we denote $N_{0}$ as the number of $0$'s and $N_{\pi}$ as the number of $\pi$'s of its three components ${\bf k}_{0}=(k_{0x},k_{0y},k_{0z})$. It is convenient to introduce the dimensionless parameter $M=\delta t/t$ and discuss the phase transition accordingly. Applying Eqs.~(\ref{FkGamma_definition}) and (\ref{scaling_procedure_general}) with ${\bf k}_{s}=(1,1,1)/\sqrt{3}$, and denoting $M^{\prime}-M=dM$, $dl=\delta k^{2}$, one obtains the RG equation
\begin{eqnarray}
\frac{dM}{dl}&=&\frac{1+M}{6}-\frac{\left(N_{0}-N_{\pi}\right)\left(1+M\right)}{3\left(1+N_{0}-N_{\pi}+M\right)}
=\beta_{dia}(M)\;,
\nonumber \\
\xi&=&\left|\frac{\beta_{dia}(M)}{1+N_{0}-N_{\pi}+M}\right|^{1/2}\; .
\label{RG_eq_diamond_lattice}
\end{eqnarray}
This RG equation correctly captures the phase transitions which occur when $M=\delta t/t=-1-N_{0}+N_{\pi}$,
corresponding to the gap-closing condition at ${\bf k}_{0}$. Since $N_{0}-N_{\pi}=\left\{-3,-1,1,3\right\}$, there are four critical points $\delta t=\left\{2t,0,-2t,-4t\right\}$ that are correctly identified by the RG flow in Fig.~\ref{fig:diamond_lattice_model}(b). The length scale $\xi$ diverges or vanishes at the critical points, and vanishes at the fixed point, as indicated in Fig.~\ref{fig:diamond_lattice_model}(a), indicating that it is an adequate index to characterize the scale invariance at the critical points and the fixed points\cite{Huang87}.




\section{Scaling scheme using second derivative of the Pfaffian}

Next, we consider the scaling scheme which uses the second derivative of the Pfaffian.
As before, we first discuss the general formalism and then illustrate the RG scheme by
use of examples.

\subsection{General formalism}

Let us consider topological systems whose ${\mathbb Z}_{2}$ invariants are calculated by 
the sign of the Pfaffian at high-symmetry points\cite{Fu07,Fu07_2}
\begin{eqnarray}
\left(-1\right)^{\nu}=\prod_{i}{\rm Sgn}\left({\rm Pf}[m({\bf k}_{0,i})]\right)\; ,
\label{Z2_index_from_sgn_Pfaffian}
\end{eqnarray}
where ${\bf k}_{0,i}$ is the $i$-th high-symmetry point. Note that at ${\bf k}_{0,i}$ the $m_{\alpha\beta}$ matrix is equal to the sewing matrix $w_{\alpha\beta}$.
 The Pfaffian ${\rm Pf}(m)$ in these models is not necessarily complex, hence it may have no phase gradient and the scaling scheme of the previous section is not applicable. For the examples below, we found that 
\begin{eqnarray}
{\rm Pf} [ m({\bf k}_{0,i}) ]
=\frac{{\rm Pf}[m({\bf k}_{0,i})]}{|{\rm Pf}[m({\bf k}_{0,i})]|}={\rm Sgn}\left({\rm Pf}[m({\bf k}_{0,i})]\right)=\pm 1\;,
\nonumber \\
\end{eqnarray}
i.e., ${\rm Pf}[ m({\bf k}_{0,i}) ]$ remains at $\pm 1$ within a particular topological phase (see Fig.~\ref{fig:BHZ_model_Pfaffian} below).

Hence, instead of using the phase of ${\rm Pf} (m)$, we propose to use the second derivative of the Pfaffian as the scaling function
\begin{eqnarray}
F({\bf k},M)=\partial_{k_{s}}^{2}{\rm Pf}[m({\bf k},M)],
\label{scaling_fn_2nd_derivative}
\end{eqnarray}
where the scaling direction ${\hat{\bf k}}_{s}\parallel{\bf k}_{0,1}-{\bf k}_{0,2}$ points from one high-symmetry point ${\bf k}_{0,1}$ to another ${\bf k}_{0,2}$. This is a well-defined choice because the 1D integral of Eq.~(\ref{scaling_fn_2nd_derivative}) along the path from ${\bf k}_{0,1}$ to ${\bf k}_{0,2}$ is a topological invariant 
\begin{eqnarray}
{\cal C}^{\prime}&=&\int_{{\bf k}_{0,1}}^{{\bf k}_{0,2}}dk\;F({\bf k},M)=\int_{{\bf k}_{0,1}}^{{\bf k}_{0,2}}dk\;\partial_{k_{s}}^{2}{\rm Pf}[m({\bf k},M)]
\nonumber \\
&=&\partial_{k_{s}}{\rm Pf}[m({\bf k},M)]|_{{\bf k}_{0,2}}-\partial_{k_{s}}{\rm Pf}[m({\bf k},M)]|_{{\bf k}_{0,1}}=0\; .
\nonumber \\
\label{C_prime}
\end{eqnarray}
That is, the integration is equal to the first derivative of ${\rm Pf}[m({\bf k},M)]$ at the high-symmetry points, which is always zero because ${\rm Pf}[m({\bf k}_{0,i},M)]=\pm 1$ is an extremum. 
Applying the scaling procedure, Eq.~(\ref{scaling_procedure_general}), with $F({\bf k},M)$ given by Eq.~\eqref{scaling_fn_2nd_derivative} then reduces the second derivative of ${\rm Pf}[m({\bf k},M)]$ without changing the fact that it is an extremum at ${\bf k}_{0,i}$. Hence, the system flows away from the critical point at which the second derivative of ${\rm Pf}[m({\bf k},M)]$ at ${\bf k}_{0,i}$ diverges (see Fig.~\ref{fig:BHZ_model_Pfaffian} below). The second derivative is reduced under this scaling procedure because, since the result of integration ${\cal C}^{\prime}$ is always zero, the scaling function itself is the deviation function $F({\bf k},M)=F_{v}({\bf k},M)$ whose value at ${\bf k}_{0,i}$ is reduced to zero as discussed after Eq.~(\ref{F_Ff_Fv}). Note that since ${\cal C}^{\prime}$ in Eq.~(\ref{C_prime}) is always zero, it is not directly related to the ${\mathbb Z}_{2}$ invariant in Eq.~(\ref{Z2_index_from_sgn_Pfaffian}). The length scale $\xi$ can also be introduced using Eq.~(\ref{correlation_length_definition}), which is again an appropriate index to characterize the scale invariance at the critical points and fixed points in this scaling scheme.

\begin{figure}[ht]
\begin{center}
\includegraphics[clip=true,width=0.99\columnwidth]{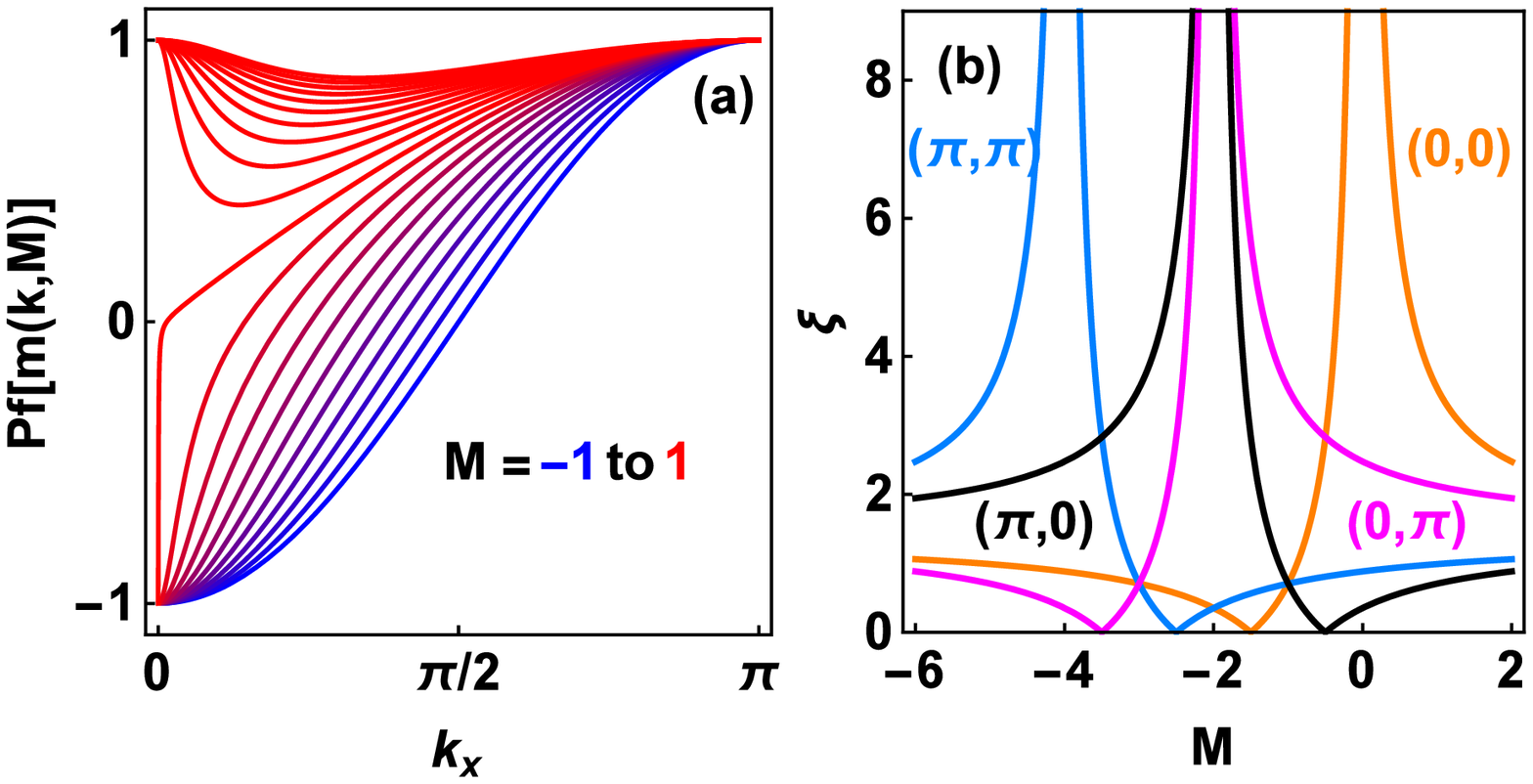}
\includegraphics[clip=true,width=0.7\columnwidth]{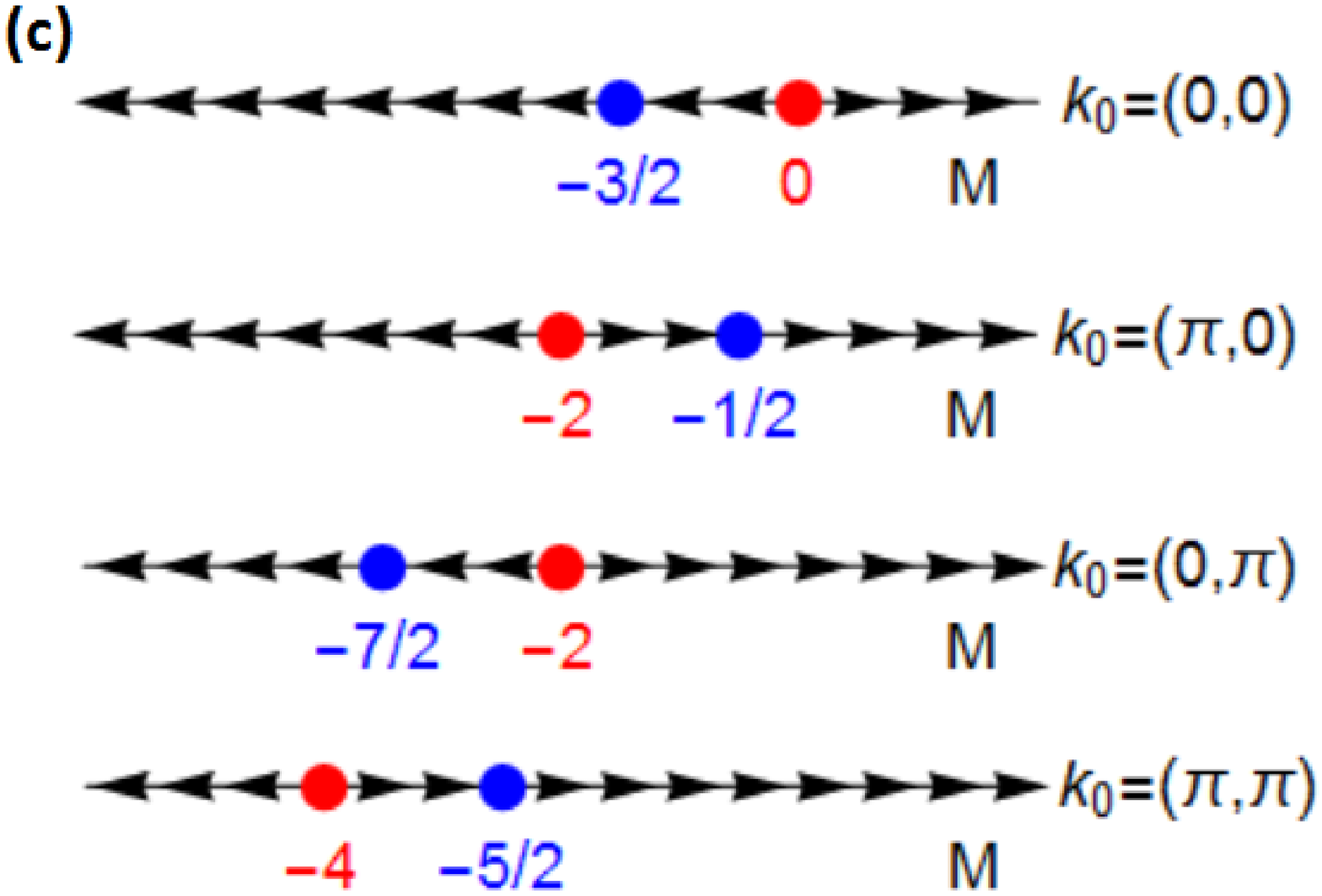}
\caption{(color online) (a) The Pfaffian ${\rm Pf}\left[m({\bf k},M)\right]$ of the BHZ model~\eqref{BHZ_Dirac_Hamiltonian} along $k_{x}$ at $k_{y}=0$. One sees that the Pfaffian remains at $\pm 1$ at ${\bf k}=(0,0)$ and $(\pi,0)$, but the second derivative $F({\bf k},M)=\partial_{k_{x}}^{2}{\rm Pf}\left[m({\bf k},M)\right]$ diverges and changes sign at ${\bf k}=(0,0)$ at the critical point $M_{c}=0$. (b) The length scale $\xi$ defined from the divergence of $F({\bf k},M)$ at each ${\bf k}_{0}$. (c) The RG flow of $M$. The fixed points $M_{f}$'s (blue dots) are stable on one side of the $M_{f}$ but unstable on the other side. The red and blue dots denote critical points and fixed points, respectively. 
} 
\label{fig:BHZ_model_Pfaffian}
\end{center}
\end{figure}

\subsection{Bernevig-Hughes-Zhang (BHZ) model}

The Pfaffian calculated from the BHZ model   of symmetry class AII  is a real function that has no phase gradient, so we use the scaling scheme based on the second derivative of the Pfaffian. We follow the notation in Ref.~\onlinecite{Bernevig13}. Consider the basis for spinful $s$ and $p$ orbitals
\begin{eqnarray}
\psi=\left(s\uparrow,p\uparrow,s\downarrow,p\downarrow\right)^{T}\;,
\end{eqnarray}
and the following representation for $\Gamma_{a}$ matrices
\begin{eqnarray}
\Gamma_{a}=\left\{\sigma_{z}\otimes s_{x},I\otimes s_{y},I\otimes s_{z},\sigma_{x}\otimes s_{x},\sigma_{y}\otimes s_{x}\right\}\; .
\end{eqnarray}
With these definitions, the Dirac Hamiltonian is expressed as 
\begin{eqnarray}
&&H({\bf k})=\sum_{a=1}^{3}d_{a}({\bf k})\Gamma_{a}
\nonumber \\
&&=\sin k_{x}\Gamma_{1}+\sin k_{y}\Gamma_{2}
+\left(2+M-\cos k_{x}-\cos k_{y}\right)\Gamma_{3} .
\nonumber \\
\label{BHZ_Dirac_Hamiltonian}
\end{eqnarray}
The four eigenenergies are $E_{\alpha}=\left\{-d,-d,d,d\right\}$, where $d=\sqrt{d_{1}^{2}+d_{2}^{2}+d_{3}^{2}}$. The matrix element of the time-reversal operator 
\begin{eqnarray}
\Theta=-i\sigma_{y}\otimes IK=\left(\begin{array}{cccc}
0 & 0 & -1 & 0  \\
0 & 0 & 0 & -1  \\
1 & 0 & 0 & 0   \\
0 & 1 & 0 & 0
\end{array}
\right)K
\label{TR_operator_BHZ}
\end{eqnarray}
for the two occupied states is
\begin{eqnarray}
m_{12}&=&\langle e_{1}|\Theta|e_{2}\rangle
=\frac{d_{3}}{d}={\rm Pf}(m)\;.
\end{eqnarray}
The quantity $d_{3}/d$ determines the topology of the system according to Eq.~(\ref{Z2_index_from_sgn_Pfaffian}). From Eq.~(\ref{scaling_fn_2nd_derivative}), without loss of generality, we choose ${\hat{\bf k}}_{s}=(1,0)={\bf k}_{x}$ to define
\begin{eqnarray}
F({\bf k},M)=\partial_{k_{x}}^{2}\left(\frac{d_{3}}{d}\right)
\label{BHZ_scaling}
\end{eqnarray}
and then apply Eq.~(\ref{scaling_procedure_general}). For the Dirac Hamiltonian in Eq.~(\ref{BHZ_Dirac_Hamiltonian}), using $dl=\delta k^{2}$ and $M^{\prime}=M+dM$, leads to the generic RG equation
\begin{eqnarray}
&&\frac{dM}{dl}=\frac{\left(M-M_{f}\right)^{2}}{M-M_{c}}\equiv\beta_{{\bf k}_{0}}(M)\;,
\nonumber \\
&&\xi=\left|\frac{2\beta_{{\bf k}_{0}}(M)}{M-M_{c}}\right|^{1/2}\; .
\end{eqnarray}
For the four choices of ${\bf k}_{0}$, one obtains 
\begin{eqnarray}
&&{\bf k}_{0}=(0,0):\;\;M_{c}=0\;,\;M_{f}=-3/2\;,
\nonumber \\
&&{\bf k}_{0}=(\pi,\pi):\;\;M_{c}=-4\;,\;M_{f}=-5/2\;,
\nonumber \\
&&{\bf k}_{0}=(\pi,0):\;\;M_{c}=-2\;,\;M_{f}=-1/2\;,
\nonumber \\
&&{\bf k}_{0}=(0,\pi):\;\;M_{c}=-2\;,\;M_{f}=-7/2\;,
\end{eqnarray}
which correctly captures the critical point at $M_{c}$ caused by gap-closing at the corresponding ${\bf k}_{0}$. The fixed points $M_{f}$ have the meaning that $F({\bf k},M)$ flattens to second order around ${\bf k}_{0}$ hence $\beta(M_{f})=0$. However, at $M\apprge M_{f}$ and $M\apprle M_{f}$ the RG equation $\beta(M)$ are of the same sign, i.e., RG flows are along the same direction, so the $M_{f}$ is an unstable fixed point on one side of $M_{f}$ and stable on the other side, as indicated in Fig.~\ref{fig:BHZ_model_Pfaffian}(c) by the blue dots.

\subsection{Topological superconductor}

The 2D topological superconductor of class DIII is another example of this scaling scheme. Starting from the $4\times 4$ Dirac matrices\cite{Ryu10}
\begin{eqnarray}
\alpha_{i}=\left(\begin{array}{cc}
0 & \sigma_{i}   \\
\sigma_{i} & 0   
\end{array}
\right)\;,\;
\beta=\left(\begin{array}{cc}
1 & 0   \\
0 & -1   
\end{array}
\right)\;,\;
M^{5}=\left(\begin{array}{cc}
0 & 1   \\
1 & 0   
\end{array}
\right)\;,
\end{eqnarray}
we consider the following continuum description of the class DIII superconductor
\begin{eqnarray}
&&H({\bf k})=k_{x}\alpha_{x}+k_{y}\alpha_{y}-iM\beta\gamma^{5}\;.
\end{eqnarray}
Denoting $\lambda=\sqrt{k_{x}^{2}+k_{y}^{2}+M^{2}}=\sqrt{k^{2}+M^{2}}$, the eigenenergies are $\left\{E_{1},E_{2},E_{3},E_{4}\right\}=\left\{-\lambda,-\lambda,\lambda,\lambda\right\}$. The time-reversal operator is\cite{Ryu10} 
\begin{eqnarray}
\Theta=\left(\begin{array}{cccc}
0 & 0 & 0 & -i  \\
0 & 0 & i & 0  \\
0 & -i & 0 & 0  \\
i & 0 & 0 & 0
\end{array}
\right)K\;.
\end{eqnarray}
whose matrix element for to the filled bands is
\begin{eqnarray}
m_{12}=\langle e_{1}|\Theta|e_{2}\rangle=M/\lambda={\rm Pf}\left[ m({\bf k})\right] \;.
\end{eqnarray}
The proper quantity to rescale is thus the second derivative of $m_{12}$,
\begin{eqnarray}
F({\bf k},M)=\partial_{k}^{2}\left(\frac{M}{\lambda}\right)\;,
\end{eqnarray}
where the scaling direction can be along any radial direction ${\hat{\bf k}}_{s}={\hat{\bf k}}$, since the problem is rotationally symmetric. Using $dl=\delta k^{2}$, $M^{\prime}=M+dM$, and the only high symmetry point ${\bf k}_{0}=(0,0)$, Eq.~(\ref{scaling_procedure_general}) gives
\begin{eqnarray}
&&\frac{dM}{dl}=\frac{9}{4M}\equiv\beta_{TSC}(M)\;,
\nonumber \\
&&\xi=\left|\frac{2\beta_{TSC}(M)}{M}\right|^{1/2}\;,
\end{eqnarray}
which correctly captures the critical point at $M_{c}=0$.

\section{Conclusions}

In summary, we show that for inversion-symmetric topological insulators and topological  superconductors characterized by ${\mathbb Z}_{2}$ topological invariants, a continuous scaling scheme can be constructed based on the Pfaffian of the time-reversal operator. Out of various possibilities for the definition of  the ${\mathbb Z}_{2}$ invariant~\cite{Chiu15}, we focus on those that use the Pfaffian of Eq.~(\ref{time_reversal_operator_matrix}). In these cases the Pfaffian is a continuous function such that it can be rescaled to judge the topological phase transitions. The scaling scheme renormalizes the scaling function, which is either the phase gradient or the second derivative of the Pfaffian depending on which scenario is more conveniently applicable for the model. 
Despite that different systems require different choices of the scaling function $F({\bf k},M)$, we found that    in general $F({\bf k},M)$ satisfies the following criterion: (1) Its integral over a certain spatial dimension $d^{\prime}$ (not necessarily equal to $d$) is a topological invariant
\begin{eqnarray}
{\cal C}=\int d^{d^{\prime}}{\bf k}\;F({\bf k},M)\;.
\end{eqnarray}
(2) It diverges at certain high-symmetry points as the tuning parameter approaches the critical point $M\rightarrow M_{c}$
\begin{eqnarray}
\lim_{M\rightarrow M_{c}}|F({\bf k}_{0},M)|=\infty\;.
\end{eqnarray}
(3) The divergence is of different sign when $M$ approaches $M_{c}$ from below or from above,
\begin{eqnarray}
\lim_{M\rightarrow M_{c}^{+}}{\rm Sgn}\left[F({\bf k}_{0},M)\right]=-\lim_{M\rightarrow M_{c}^{-}}{\rm Sgn}\left[F({\bf k}_{0},M)\right]\;.
\end{eqnarray}
The scaling procedure, Eq.~(\ref{scaling_procedure_general}), reduces the divergence of $F({\bf k}_{0},M)$ while keeping its sign.
Hence the system is gradually flowing away from the critical point $M_{c}$. This should serve as the general criterion for any system, and is expected to hold even for systems that lie  outside of the two scaling schemes addressed here (if there is any). Moreover, we found that the length scale defined from the divergence of either the phase gradient or the second derivative of the Pfaffian displays a universal critical behavior 
\begin{eqnarray} \label{universal_critical}
\xi\propto\left|M-M_{c}\right|^{-1} .
\end{eqnarray}
This critical behavior is shared by all the models that we examined, which covers a wide range of dimensions and symmetry classes.

Finally, we remark that an important question for future studies is whether  scaling schemes are also applicable to \emph{interacting} topological systems. Furthermore, one could consider scaling schemes that renormalize not the Hamiltonian
but the entanglement spectrum, which shows divergences at topological phase transitions\cite{ryu_hatsugai_06}. The clarification of these questions requires further investigations. 



\begin{thebibliography}{99}


\bibitem{Kitaev01}
Kitaev A V 2001 Phys. Usp. {\bf 44} 131 

\bibitem{Lutchyn10}
Lutchyn R M, Sau J D, and Das Sarma S 2010 Phys. Rev. Lett. {\bf 105} 077001 

\bibitem{Oreg10}
Oreg Y, Refael G, and von Oppen F 2010 Phys. Rev. Lett. {\bf 105} 177002 

\bibitem{Mourik12}
Mourik V, Zuo K, Frolov S M, Plissard S R, Bakkers E P A M, and Kouwenhoven L P 2012 Science {\bf 336} 1003 

\bibitem{Su79}
Su W P, Schrieffer J R, and Heeger A J 1979 Phys. Rev. Lett. {\bf 42} 1698

\bibitem{Choy11}
Choy T-P, Edge J M, Akhmerov A R, and Beenakker C W J 2011 Phys. Rev. B {\bf 84} 195442 

\bibitem{NadjPerge13}
Nadj-Perge S, Drozdov I K, Bernevig B A, and Yazdani A 2013 Phys. Rev. B {\bf 88} 020407 

\bibitem{Braunecker13}
Braunecker B and Simon P 2013 Phys. Rev. Lett. {\bf 111} 147202 

\bibitem{Pientka13}
Pientka F, Glazman L I, and von Oppen F 2013 Phys. Rev. B {\bf 88} 155420 

\bibitem{Klinovaja13}
Klinovaja J, Stano P, Yazdani A, and Loss D 2013 Phys. Rev. Lett. {\bf 111} 186805 

\bibitem{Vazifeh13}
Vazifeh M M and Franz M 2013 Phys. Rev. Lett. {\bf 111} 206802 

\bibitem{Rontynen14}
R\"{o}ntynen J and Ojanen T 2014 Phys. Rev. B {\bf 90} 180503 

\bibitem{Kim14}
Kim Y, Cheng M, Bauer B, Lutchyn R M, and Das Sarma S 2014 Phys. Rev. B {\bf 90} 060401 

\bibitem{Sedlmayr15}
Sedlmayr N, Aguiar-Hualde J M, and Bena C 2015 Phys. Rev. B {\bf 91} 115415

\bibitem{Chen15}
Chen W and Schnyder A P 2015 Phys. Rev. B {\bf 92} 214502 


\bibitem{Schnyder08}
Schnyder A P, Ryu S, Furusaki A, and Ludwig A W W 2008 Phys. Rev. B {\bf 78} 195125 

\bibitem{Kitaev09}
Kitaev A 2009 AIP Conf. Proc. {\bf 1134} 22 

\bibitem{Chiu15}
Chiu C-K, Teo J C Y, Schnyder A P, and Ryu S, arXiv:1505.03535.


\bibitem{Chen16}
Chen W 2016 J. Phys.: Condens. Matter {\bf 28} 055601 


\bibitem{Kadanoff66}
Kadanoff L P 1966 Physics {\bf 2} 263 


\bibitem{Berry84}
Berry M V 1984 Proc. R. Soc. London, Ser. A {\bf 392} 45 

\bibitem{Xiao10}
Xiao D, Chang M-C, and Niu Q 2010 Rev. Mod. Phys. {\bf 82} 1959 

\bibitem{Thouless82}
Thouless D J, Kohmoto M, Nightingale M P, and den Nijs M 1982 Phys. Rev. Lett. {\bf 49} 405 

\bibitem{Zak89}
Zak J 1989 Phys. Rev. Lett. {\bf 62} 2747 


\bibitem{Kane05}
Kane C L and Mele E J 2005 Phys. Rev. Lett. {\bf 95} 146802 

\bibitem{Fu07}
Fu L and Kane C L 2007 Phys. Rev. B {\bf 76} 045302 

\bibitem{Fu07_2}
Fu L, Kane C L, and Mele E J 2007 Phys. Rev. Lett. {\bf 98} 106803 

\bibitem{Moore07}
Moore J E and Balents L 2007 Phys. Rev. B {\bf 75} 121306 


\bibitem{Fu06}
Fu L and Kane C L 2006 Phys. Rev. B {\bf 74} 195312 




\bibitem{Chen04}
Chen W, Chang M-S, Lin H-H, Chang D, and Mou C-Y 2004 Phys. Rev. B {\bf 70} 205413 

\bibitem{Chang05}
Chang M-S, Chen W, Lin H-H 2005 Prog. Theor. Phys. Supp. {\bf 160} 79 



\bibitem{Fruchart13}
Fruchart M and Carpentier D 2013 C. R. Physique {\bf 14} 779 










\bibitem{Bernevig06}
Bernevig B A, Hughes T L, and Zhang S-C 2006 Science {\bf 314} 1757 



















\bibitem{Huang87}
Huang K 1987 {\it Statistical Mechanics} John Wiley and Sons, p. 445.

\bibitem{Bernevig13}
Bernevig B A with Hughes T L 2013 {\it Topological Insulators and Topological Superconductors}, Princeton University Press, Ch. 8, 9, 13, 16, and 17.









\bibitem{Ryu10}
Ryu S, Schnyder A P, Furusaki A, and Ludwig A W W 2010 New J. Phys. {\bf 12} 065010 

\bibitem{ryu_hatsugai_06}
Ryu S and Hatsugai Y 2006 Phys. Rev. B {\bf 73} 245115 


\end{thebibliography}
\end{document}